\documentclass[aps,jcp,twocolumn,groupedaddress]{revtex4-1}
\usepackage{graphicx,amsmath,color}
\begin{document}

\title{Isostaticity and the solidification of semiflexible polymer melts}
\author{Christian O. Plaza-Rivera}
\author{Hong T. Nguyen}
\author{Robert S. Hoy}
\email{rshoy@usf.edu}
\affiliation{Department of Physics, University of South Florida, Tampa, FL 33620, USA}
\date{\today}
\begin{abstract}
Using molecular dynamics simulations of a tangent-soft-sphere bead-spring polymer model, we examine the degree to which semiflexible polymer melts solidify at isostaticity.  
Flexible and stiff chains crystallize when they are isostatic as defined by appropriate degree-of-freedom-counting arguments.
Semiflexible chains also solidify when isostatic if a generalized isostaticity criterion that accounts for the slow freezing out of configurational freedom as chain stiffness increases is employed.
The dependence of the average coordination number at solidification $Z(T_s)$ on chains' characteristic ratio $C_\infty$ has the same functional form [$Z \simeq a - b\ln(C_\infty)$] as the dependence of the average coordination number at jamming $Z(\phi_J)$ on $C_\infty$ in athermal systems, suggesting that jamming-related phenomena play a significant role in thermal polymer solidification.
\end{abstract}
\maketitle

\section{Introduction}

Traditional analytic criteria for solidification, such as those based on classical nucleation theory, typically work very poorly both for liquids with strong glassforming tendency and for polymeric liquids.
This failure creates a need for alternative criteria predicting these systems' solidification transitions.
Several have been proposed, such as the splitting of the first peak in the pair correlation function $g(r)$, the height of this peak $g_{max}$ and its ratio to the value $g_{min}$ of $g(r)$ at its first minimum, and other criteria based on $g(r)$ or local, cluster-level structure.\cite{wendt78,vanblaaderen95,corwin05,caswell13,royall08,royall15}
Criteria based on the average coordination number $\left<Z\right>$, such as the famous result that systems of spherical particles jam at isostaticity [$\left<Z\right> = 2d$, where $d$ is spatial dimension]\cite{maxwell64,ohern02} naturally fit into both of these categories.

Recent work\cite{karayiannis08,hoy12,lappala16,zaccone13} has suggested an interesting connection between isostaticity and solidification of polymeric liquids: that solidification occurs when the average number of noncovalent contacts per monomer $\left<Z_{nc}\right>$ exceeds its isostatic value $\left<Z_{nc}^{iso}\right>$.
These studies focused on liquids of fully flexible\cite{karayiannis08,hoy12,lappala16,zaccone13} or infinitely stiff\cite{zaccone13} chains, for which the definition of  $\left<Z_{nc}^{iso}\right>$ is straightforward.
However, \textit{finite} chain stiffness is well known to strongly and nontrivially affect polymer soldification in both thermal\cite{ballauff89,strobl07,nguyen15} and athermal\cite{zou09,hoy17} systems.
Here we examine the connection of isostaticity to polymer solidification using molecular dynamics simulations of a simple crystallizable bead-spring model\cite{hoy13} with continuously variable chain stiffness.
By considering chains ranging from flexible to rodlike and employing a suitably generalized isostaticity criterion, we show that these model  polymeric liquids are very generally isostatic at their solidification temperatures.

Consider $d=3$ chains of length $N$ with monomer positions $\vec{r}_i$, covalent bond lengths $\ell_i = |\vec{r}_{i+1}-\vec{r}_i|$, and bond angles $\theta_i = \cos^{-1}[\vec{\ell}_{i-1}\cdot\vec{\ell}_{i}/(\ell_{i-1} \ell_{i})]$.
Maxwell's isostaticity criterion\cite{maxwell64} can be written as $\left< Z^{iso} \right> = 2N^{-1}(Nd- n_{constr})$, where $n_{constr}$ is the number of holonomic constraints per chain.
Fixed-length  ($\ell = \ell_0$) covalent bonds and fixed bond angles ($\theta=\theta_0$) respectively supply $N-1$ and $N-2$ constraints per chain.\cite{phillips79,thorpe83}
Fully flexible chains with fixed-length covalent bonds have\cite{zaccone13}
\begin{equation}
\left<Z_{nc}^{iso}\right> = \left<Z_{nc}^{iso}\right>_{flex} = \displaystyle\frac{2[3N - (N-1)]}{N} = 4 + \displaystyle\frac{2}{N},
\label{eq:isoflex}
\end{equation}
while infinitely stiff chains that also have fixed bond angles have\cite{zaccone13} 
\begin{equation}
\left<Z_{nc}^{iso}\right> = \left<Z_{nc}^{iso}\right>_{stiff} =  \displaystyle\frac{2[3N-(N-1)-(N-2)]}{N} = 2 + \displaystyle\frac{6}{N}.
\label{eq:isorigid}
\end{equation}
For semiflexible chains, a more general isostaticity criterion intermediate between Eqs.\ \ref{eq:isoflex} and \ref{eq:isorigid} may apply.\cite{micoulaut16}
If isostaticity controls solidification but angular degrees of freedom are \textit{gradually} frozen out as chain stiffness increases, the average number of noncovalent contacts per monomer at the solidifcation temperature $T_s$, $\left<Z_{nc}(T_s)\right>$, should vary smoothly from $\left<Z_{nc}^{iso}\right>_{flex}$ to $\left<Z_{nc}^{iso}\right>_{stiff}$.
Below, we use molecular dynamics simulations to show that this indeed occurs in model systems, and derive a generalized isostaticity criterion describing the phenomenon.

\section{Model and Methods}

Our simulations employ the soft-pearl-necklace polymer model described at length in Refs.\cite{hoy13,nguyen15}
It is comparable to the Kremer-Grest bead-spring model,\cite{kremer90} but possesses crystalline ground states.
All monomers have mass $m$ and interact via the truncated and shifted Lennard-Jones potential
\begin{equation}
U_{LJ}(r) = \epsilon\left[\left(\displaystyle\frac{\sigma}{r}\right)^{12} - \left(\displaystyle\frac{\sigma}{r_c}\right)^{12} - 2\left(\left(\displaystyle\frac{\sigma}{r}\right)^{6} - \left(\displaystyle\frac{\sigma}{r_c}\right)^{6}\right)\right],
\label{eq:LJpot}
\end{equation}
where $\epsilon$ is the intermonomer binding energy and $r_c = 2^{7/6}\sigma$ is the cutoff radius.
Bonds between adjacent beads along the chain backbone are modeled using the harmonic potential
\begin{equation}
U_c(\ell) = \displaystyle\frac{k_c}{2}\left(\ell-\sigma\right)^2,
\label{eq:Ubond}
\end{equation}
where $\ell$ is bond length and $k_c$ is the bond stiffness.
The large value of $k_c$ employed here ($600\epsilon/\sigma^2$) produces bonds of nearly fixed $\ell$; $U_c(\ell)$ effectively acts as a holonomic constraint fixing $\ell= \ell_0 = \sigma$ and preventing chain crossing.\cite{hoy13}    
Bending stiffness is included using the standard potential\cite{auhl03}
\begin{equation}
U_b(\theta) = k_{bend}(1 - cos(\theta)),
\label{eq:Ubend}
\end{equation}
which favors straight trimers (sets the equilibrium bond angle $\theta_0 = 0$).
Fully flexible chains have $k_{bend} = 0$, and rigid-rod-like chains are obtained in the limit $k_{bend} \to \infty$.
Here we study systems with $0 \leq k_{bend} \leq 30\epsilon$.
As detailed in Ref.\cite{nguyen15}, the model's solid morphologies -- formed by cooling from the isotropic liquid state -- range from random-walk close-packed crystals to glasses to nematic close-packed crystals over this range of $k_{bend}$.
Since its soldification dynamics\cite{nguyen16} also vary strongly with $k_{bend}$, the model is suitable for studying connections between solidification and isostaticity in a very general way.

All systems are composed of $N_{ch}=500$ chains of $N=25$ monomers.
These chains are unentangled.
Periodic boundaries are applied along all three directions of cubic simulation cells.
Systems are first thoroughly equilibrated \cite{auhl03} at temperatures well above their $k_{bend}$-dependent solidification temperatures,\cite{nguyen15}
then slowly cooled at zero pressure to $T=0$ at a rate $|\dot{T}|=10^{-6}/\tau$.
This $|\dot{T}|$ is sufficiently low to be in a limit where finite-cooling-rate effects on melt structure are small.\cite{nguyen16}
Pressure is controlled using a Nose-Hoover barostat.
The MD timestep used here is $\delta t = \tau/200$, where $\tau$ is the Lennard-Jones time unit $\sqrt{m\sigma^2/\epsilon}$.
All simulations are performed using LAMMPS.\cite{plimpton95}

\begin{figure}[htbp]
\centering
\includegraphics[width=3in]{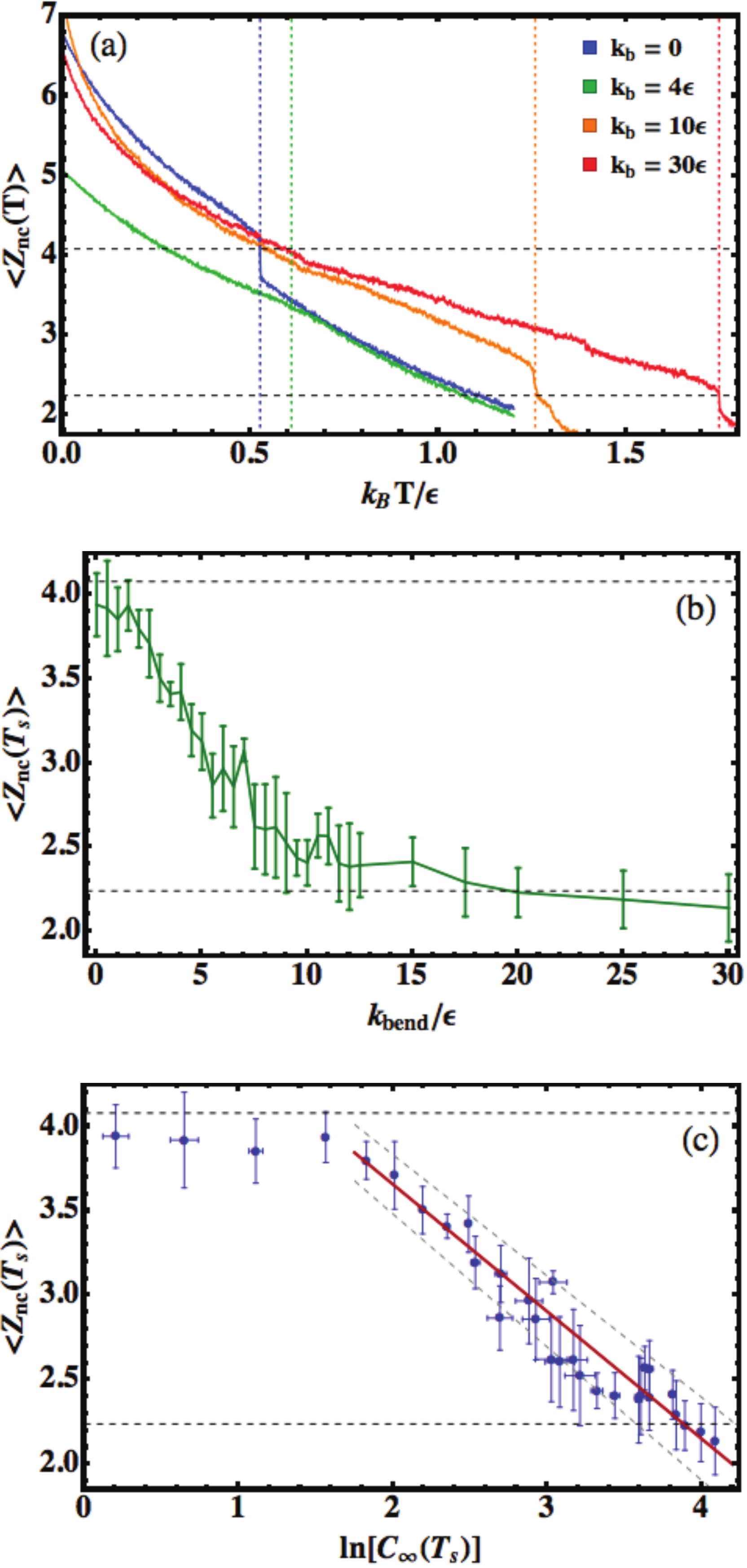}
\caption{
Measures of noncovalent repulsive contacts in slowly cooled semifexible polymer melts.  Panel (a) shows $\left< Z_{nc}(T) \right>$: blue, green, orange and red curves show data for selected $k_{bend}$, while the correspondingly colored vertical dotted lines indicate the respective $T_s(k_{bend})$ [Table \ref{tab:Ts}].
Panel (b) shows the $k_{bend}$-dependence of $\left< Z_{nc}(T_s) \right>$ for all systems.  
In panel (c), points show the same  $\left< Z_{nc}(T_s) \right>$ as in panel (c), but plotted vs.\ $C_{\infty}(T_s)$.
The red solid line shows a fit to our generalized isostaticity criterion (Eq.\ \ref{eq:genisocrit}), and the angled dotted lines show the maximal statistical uncertainties on this fit, including uncertainties on both the slope $b$ and the intercept $\left[\left<Z_{nc}^{iso}\right>_{flex} - \left<Z_{nc}^{iso}\right>_{stiff}\right]/b$. 
In all panels, the upper and lower horizontal gray dotted lines respectively indicate $\left<Z_{nc}^{iso}\right>_{flex} = 4.08$ and $\left<Z_{nc}^{iso}\right>_{stiff} = 2.24$ for these $N=25$ systems.  Results from separate studies of $N= 13$ and $N=50$ systems for selected $k_{bend}$ (not shown) were in all ways consistent with those shown here.}
\label{fig:1}
\end{figure}

\section{Results}

Ref.\cite{nguyen15} presented a detailed analysis of these systems' solidification behavior for $k_{bend} \leq 12.5\epsilon$, but did not consider staticity.
Figure \ref{fig:1} presents staticity-related results.
Panel (a) shows how $\left< Z_{nc}(T) \right>$ increases during cooling for four representative chain stiffnesses: flexible ($k_{bend} = 0$), semiflexible ($k_{bend} = 4\epsilon$), semistiff ($k_{bend} = 10\epsilon$), and stiff ($k_{bend} = 30\epsilon$).
Here
\begin{equation}
\left< Z_{nc} \right> = \displaystyle\frac{2}{N_{ch}N} \displaystyle\sum_{i=1}^{N_{ch}N} \displaystyle\sum_{j=i+1}^{N_{ch}N} \Theta(1-r_{ij}) f_{ij},
\label{eq:zncdef}
\end{equation}
where $\Theta(x)$ is the Heaviside step function; $f_{ij} = 0$ if monomers $i$ and $j$ are covalently bonded and $1$ otherwise.
Thus $\left< Z_{nc} \right>$ only counts \textit{repulsively} interacting particles (those with $r_{ij} \equiv |\vec{r}_j -\vec{r}_i| < 1$) as being in contact, as is appropriate for thermal systems.\cite{micoulaut16}
Flexible chains crystallize into a random-walk close-packed (RWCP) structure wherein monomers close-pack but chains retain random-walk-like structure and are isotropically oriented.\cite{nguyen15}
Semiflexible chains form glasses; $k_{bend}=4\epsilon$ systems have been shown to be typical fragile glassformers.\cite{nguyen16}
Semistiff chains form moderately defective nematic close-packed (NCP) crystals,\cite{nguyen15} while stiff chains form nearly perfect NCP crystals.
Solidification temperatures $T_s$ increase by more than a factor of three (from $k_B T_s/\epsilon = 0.53$ for flexible chains to $k_B T_s/\epsilon = 1.75$ for stiff chains) as stiffness increases (Table \ref{tab:Ts}).
The densities of these systems at $T_s$ also drop sharply with increasing $k_{bend}$ over the same range; Table \ref{tab:Ts} reports the packing fractions $\phi_s = \phi(T_s)$, where $\phi = \pi \rho/6$ is the usual packing fraction for spherical particles [and $\rho = N_{ch}N/V$ is the monomer number density].
Thus these systems collectively exhibit a wide range of solidification behaviors.

\begin{table}
\small
\caption{Solidification temperatures, densities, and solid morphologies for selected chain stiffnesses.  Values of $T_s$ and $\phi_s$ for $k_{bend} \leq 12.5\epsilon$ and morphology descriptions were reported in Ref.\cite{nguyen15}.}
  \begin{tabular*}{0.5\textwidth}{@{\extracolsep{\fill}}lccc}
\hline
$k_{bend}/\epsilon$ & $k_B T_s/\epsilon$ & $\phi_s$ & Morphology\\
\hline
0 & 0.53 & 0.684 & RWCP\\
2 & 0.49 & 0.673 & glass/RWCP\\
4 & 0.61 & 0.646 & glass\\
6 & 0.91 & 0.606 & nematic glass\\
8 & 1.13 & 0.581 & multidomain NCP\\
10 & 1.26 & 0.580 & defected NCP\\
15 & 1.32 & 0.582 & NCP\\
20 & 1.57 & 0.558 & NCP\\
25 & 1.66 & 0.556 & NCP\\
30 & 1.75 & 0.552 & NCP\\
\hline
\end{tabular*}
\label{tab:Ts}
\end{table}

Crystallizing systems exhibit sharp, first-order-transition-like jumps in $\left< Z_{nc}(T) \right>$ at $T=T_s$.
Glassforming systems exhibit smoothly increasing $\left< Z_{nc}(T) \right>$ as $T$ decreases, with only slight cusps [discontinuities in $\partial^2 \left< Z_{nc}(T) \right>/\partial T^2$] at $T=T_s$.\footnote[1]{As in Ref.\ \cite{nguyen15}, values of $T_s$ were determined by locating the jump in packing fraction $\phi(T)$ for crystallizing systems, or the intersection of low-$T$ and high-$T$ linear fits to $\phi(T)$ for glassforming systems.}
Below $T_s$, $\left< Z_{nc}(T) \right>$ continues to increase as cooling proceeds, not because of any major structural rearrangements, but simply because systems continue to densify. 
It is clear that both flexible-chain systems and stiff-chain systems are approximately isostatic at $T = T_s$, i.e.\ they respectively have $\left< Z_{nc}(T_s) \right> \simeq \left<Z_{nc}^{iso}\right>_{flex}$ and  $\left< Z_{nc}(T_s) \right> \simeq \left<Z_{nc}^{iso}\right>_{stiff}$.
It is also clear that intermediate-stiffness systems display intermediate solidification behavior that exhibits a smooth crossover between the flexible-chain and stiff-chain limits.

 Refs.\ \cite{hoy12,lappala16} reported $\left< Z_{nc}(T_s) \right> \simeq \left<Z_{nc}^{iso}\right>_{flex}$ in single-flexible-chain systems.
Refs.\ \cite{zaccone13,lappala16} argued that $\left< Z_{nc}(T_s) \right>$ should be $\left<Z_{nc}^{iso}\right>_{flex}$ in bulk glassforming polymeric liquids when chains are fully flexible, and $\left<Z_{nc}^{iso}\right>_{stiff}$ when chains are infinitely stiff (have holonomic $\theta = \theta_0$ constraints).
Panel (b) shows $\left< Z_{nc}(T_s) \right>$ for all systems as a function of $k_{bend}$.
$\left< Z_{nc}(T_s) \right>$ is roughly constant for $k_{bend} <\sim \epsilon$, then drops sharply with increasing $k_{bend}$ until the stiff-chain $\left< Z_{nc}(T_s) \right> \simeq \left<Z_{nc}^{iso}\right>_{stiff}$ limit is approached as $k_{bend}$ exceeds $\sim 10\epsilon$.
The data in panels (a-b) clearly show that $\left< Z_{nc}(T_s) \right> \simeq \left<Z_{nc}^{iso}\right>_{flex}$ also holds true for crystal-forming flexible-chain liquids, and that $\left< Z_{nc}(T_s) \right> \simeq \left<Z_{nc}^{iso}\right>_{stiff}$ holds for crystal-forming stiff-chain liquids.
They also strongly suggest semiflexible chains' solidification behavior should be describable by a suitably generalized criterion for  $\left< Z_{nc}(T_s) \right>$.

One commonly used measure of chain stiffness that is easily connected to the configurational freedom associated with bond angles is the characteristic ratio $C_\infty = (1+\left<\cos(\theta)\right>)/(1-\left<\cos(\theta)\right>)$.
$C_\infty = 1$ for ideally flexible chains with no excluded volume, $\sim 1.7$ for $k_{bend} = 0$ chains,\cite{auhl03,foteinopoulou08} and $\infty$ for $k_{bend} = \infty$ rod-like chains.
Thus the variation of $C_\infty$ can be taken as a rough proxy for the slow freezing out of the bond-angular degrees of freedom as chain stiffness increases and/or temperature decreases.
While many-body effects can considerably alter $C_\infty$ in bulk systems (e.g.\ dense liquids at $T=T_s$ or athermal systems at $\phi = \phi_J$\cite{foteinopoulou08,hoy17}), it is still reasonable to posit that a generalized isostaticity criterion based on $C_\infty$ exists.
One postulate is that the \textit{effective} number of holonomic constraints per chain at solidification is
\begin{equation}
n_{constr}^{eff}[C_\infty(T_s)] = (N-1) + (N-2) g\left[C_\infty(T_s)\right],
\label{eq:nconst}
\end{equation}
where $g(C_\infty)$ smoothly increases from $0$ to $1$ as $C_{\infty}$ varies from $1$ to $\infty$.
Then a potential generalized isostaticity criterion is
\begin{equation}
\left< Z_{nc}^{iso} \right>_{gen} = 2N^{-1}(Nd- n_{constr}^{eff}[C_\infty(T_s)]).
\label{eq:zisogen}
\end{equation}
with $n_{constr}^{eff}$ given by Eq.\ \ref{eq:nconst}.
This formula automatically satifsfies $\left< Z_{nc}^{iso} \right>_{gen} = \left< Z_{nc}^{iso} \right>_{flex}$ when $C_\infty = 1$ and $\left< Z_{nc}^{iso} \right>_{gen} = \left< Z_{nc}^{iso} \right>_{flex}$ when $C_\infty = \infty$, and thus is consistent with Eqs.\ \ref{eq:isoflex} and \ref{eq:isorigid}.
Since it is not clear how to calculate $g(C_\infty)$ ``ab initio'', we will attempt to determine a functional form for $g(C_\infty)$  by examining our simulation-generated dataset.

Figure \ref{fig:1}(c) shows $\left< Z_{nc}(T_s) \right>$ for all systems as a function of $C_{\infty}(T_s)$.
Systems with $k_{bend}/\epsilon <\sim 1.5$ [$C_\infty(T_s) <\sim 7$] are apparently in the flexible-chain limit where $g(C_\infty) \simeq 0$ and $\left< Z_{nc}(T_s) \right> \simeq \left<Z_{nc}^{iso}\right>_{flex}$.
For larger $k_{bend}$ and $C_\infty(T_s)$, the decrease of $\left< Z_{nc}(T_s) \right>$ with increasing chain stiffness is approximately logarithmic in $C_\infty(T_s)$.
Stiff chains with $C_\infty(T_s) \simeq 50$ have $\left< Z_{nc}(T_s) \right> \simeq \left<Z_{nc}^{iso}\right>_{stiff}$.
The data suggest $g(C_\infty) \sim \ln[C_\infty]$ for $7 <\sim C_\infty <\sim 50$, and that a generalized isostaticity criterion of form
\begin{equation}
\left< Z_{nc}^{iso} \right>_{gen}  = \rm{min}\left[\left<Z_{nc}^{iso}\right>_{flex}, \left<Z_{nc}^{iso}\right>_{stiff} - b\ln\left( \displaystyle\frac{C_\infty}{C_\infty^{max}} \right) \right]
\label{eq:genisocrit}
\end{equation}
describes the solidification of semiflexible polymers over the full range of $C_\infty$ considered here; as shown in panel (c), the fit of Eq.\ \ref{eq:genisocrit} to the data for $k_{bend}/\epsilon \geq 2\epsilon$ is very good.
Note that this range of $C_\infty$ is comparable to the range exhibited by natural polymers, from very flexible ones such as polyethylene to stiff ones such as actin.\cite{ballauff89,strobl07}
Very stiff chains with $C_{\infty} \geq C_\infty^{max} \simeq 10^2$ lie in a different regime where chains behave as though they were single rigid-rod-like particles rather than polymers,\cite{kyrylyuk11,broedersz11} and are not considered here.

The crossover from the flexible to the semiflexible regime (i.e.\ the crossover between the two functional forms for $\left< Z_{nc}^{iso} \right>_{gen}$ given in Eq.\ \ref{eq:genisocrit}) is a subtle issue.  
The data in Fig.\ \ref{fig:1} actually suggests that polymer melts are \textit{very} slightly hypostatic at solidification, to a degree that is nearly independent of chain stiffness.
This slight deviation may be related to solidification occurring when iso/hyper-static clusters percolate rather than when $\left< Z_{nc} \right> =  \left< Z_{nc}^{iso} \right>_{gen}$,\cite{lois08} but analyses of such clusters in our systems were inconclusive.
Alternatively, the deviation may be related to thermal effects including nonperturbative effects of attractive interactions and the shape of the repulsive part of the potential,\cite{berthier09} or many-body phenomena including dimer-interlocking.\cite{schreck10}
Such effects are usually subtle and would require intensive analyses that are beyond our present scope.
Thus the generalized isostaticity criterion developed here (Eq.\ \ref{eq:genisocrit}) can be considered a peer of those proposed in Refs.\ \cite{royall08,royall15,wendt78,vanblaaderen95,corwin05,caswell13} in the sense that while it is neither rigorous nor precise, it can serve as a useful guide.

\section{Discussion and conclusions}

The trends illustrated in Fig.\ \ref{fig:1} strongly suggest that isostaticity is a broadly important concept for improving our understanding of semiflexible polymer solidification.
The $\ln(C_\infty)$ dependence of $\left< Z_{nc}(T_s) \right>$ in our thermal systems is also observed in jamming of athermal semiflexible polymers, which have $\left< Z_{nc} (\phi_J) \right> \simeq a - b\ln(C_\infty)$ in the range $10^1 <\sim C_\infty <\sim10^2$.\cite{hoy17}
This similar functional dependence of monomer coordination at solidification upon $C_\infty$ is present despite the fact that Ref.\ \cite{hoy17} employed a different angular potential [$U_b = (k_{bend}/2)(\theta - \theta_0)^2$] and varied $C_\infty$ by varying $\theta_0$ rather than $k_{bend}$.  
The common behavior supports previous work (e.g.\ Refs.\cite{zou09,karayiannis08,zaccone13,lappala16}) suggesting that jamming-related phenomena play a role in controlling polymer melt solidification despite the fact that polymer melts are highly thermal.
For example, the well-known increase in $T_g$ with increasing $C_\infty$ in microscopic synthetic polymers,\cite{strobl07,ballauff89} the observed decrease in $\phi_J$ with $C_\infty$ in athermal polymers,\cite{hoy17,zou09} and the data presented herein all form a consistent picture if one accepts the idea that all these trends are dominated by the gradual freezing out of configurational freedom as chain stiffness increases.
In conclusion, the accumulated evidence now strongly suggests that $C_\infty$ is an axis on the polymeric counterpart of Liu and Nagel's jamming-glass phase diagram.\cite{liu98}

\section{Acknowledgements}

Alessio Zaccone provided helpful discussions.
This material is based upon work supported by the National Science Foundation under Grant Nos.\ DMR-1555242 and DMR-1560090.


\begin{thebibliography}{33}%
\makeatletter
\providecommand \@ifxundefined [1]{%
 \@ifx{#1\undefined}
}%
\providecommand \@ifnum [1]{%
 \ifnum #1\expandafter \@firstoftwo
 \else \expandafter \@secondoftwo
 \fi
}%
\providecommand \@ifx [1]{%
 \ifx #1\expandafter \@firstoftwo
 \else \expandafter \@secondoftwo
 \fi
}%
\providecommand \natexlab [1]{#1}%
\providecommand \enquote  [1]{``#1''}%
\providecommand \bibnamefont  [1]{#1}%
\providecommand \bibfnamefont [1]{#1}%
\providecommand \citenamefont [1]{#1}%
\providecommand \href@noop [0]{\@secondoftwo}%
\providecommand \href [0]{\begingroup \@sanitize@url \@href}%
\providecommand \@href[1]{\@@startlink{#1}\@@href}%
\providecommand \@@href[1]{\endgroup#1\@@endlink}%
\providecommand \@sanitize@url [0]{\catcode `\\12\catcode `\$12\catcode
  `\&12\catcode `\#12\catcode `\^12\catcode `\_12\catcode `\%12\relax}%
\providecommand \@@startlink[1]{}%
\providecommand \@@endlink[0]{}%
\providecommand \url  [0]{\begingroup\@sanitize@url \@url }%
\providecommand \@url [1]{\endgroup\@href {#1}{\urlprefix }}%
\providecommand \urlprefix  [0]{URL }%
\providecommand \Eprint [0]{\href }%
\providecommand \doibase [0]{http://dx.doi.org/}%
\providecommand \selectlanguage [0]{\@gobble}%
\providecommand \bibinfo  [0]{\@secondoftwo}%
\providecommand \bibfield  [0]{\@secondoftwo}%
\providecommand \translation [1]{[#1]}%
\providecommand \BibitemOpen [0]{}%
\providecommand \bibitemStop [0]{}%
\providecommand \bibitemNoStop [0]{.\EOS\space}%
\providecommand \EOS [0]{\spacefactor3000\relax}%
\providecommand \BibitemShut  [1]{\csname bibitem#1\endcsname}%
\let\auto@bib@innerbib\@empty
\bibitem [{\citenamefont {Wendt}\ and\ \citenamefont
  {Abraham}(1978)}]{wendt78}%
  \BibitemOpen
  \bibfield  {author} {\bibinfo {author} {\bibfnamefont {H.~R.}\ \bibnamefont
  {Wendt}}\ and\ \bibinfo {author} {\bibfnamefont {F.~F.}\ \bibnamefont
  {Abraham}},\ }\href@noop {} {\bibfield  {journal} {\bibinfo  {journal} {Phys.
  Rev. Lett.}\ }\textbf {\bibinfo {volume} {41}},\ \bibinfo {pages} {1244}
  (\bibinfo {year} {1978})}\BibitemShut {NoStop}%
\bibitem [{\citenamefont {{van Blaaderen}}\ and\ \citenamefont
  {Wiltzius}(1995)}]{vanblaaderen95}%
  \BibitemOpen
  \bibfield  {author} {\bibinfo {author} {\bibfnamefont {A.}~\bibnamefont {{van
  Blaaderen}}}\ and\ \bibinfo {author} {\bibfnamefont {P.}~\bibnamefont
  {Wiltzius}},\ }\href@noop {} {\bibfield  {journal} {\bibinfo  {journal}
  {Science}\ }\textbf {\bibinfo {volume} {270}},\ \bibinfo {pages} {1177}
  (\bibinfo {year} {1995})}\BibitemShut {NoStop}%
\bibitem [{\citenamefont {Corwin}\ \emph {et~al.}(2005)\citenamefont {Corwin},
  \citenamefont {Jaeger},\ and\ \citenamefont {Nagel}}]{corwin05}%
  \BibitemOpen
  \bibfield  {author} {\bibinfo {author} {\bibfnamefont {E.}~\bibnamefont
  {Corwin}}, \bibinfo {author} {\bibfnamefont {H.~M.}\ \bibnamefont {Jaeger}},
  \ and\ \bibinfo {author} {\bibfnamefont {S.~R.}\ \bibnamefont {Nagel}},\
  }\href@noop {} {\bibfield  {journal} {\bibinfo  {journal} {Nature}\ }\textbf
  {\bibinfo {volume} {435}},\ \bibinfo {pages} {1075} (\bibinfo {year}
  {2005})}\BibitemShut {NoStop}%
\bibitem [{\citenamefont {Caswell}\ \emph {et~al.}(2013)\citenamefont
  {Caswell}, \citenamefont {Zhang}, \citenamefont {Gardel},\ and\ \citenamefont
  {Nagel}}]{caswell13}%
  \BibitemOpen
  \bibfield  {author} {\bibinfo {author} {\bibfnamefont {T.~A.}\ \bibnamefont
  {Caswell}}, \bibinfo {author} {\bibfnamefont {Z.}~\bibnamefont {Zhang}},
  \bibinfo {author} {\bibfnamefont {M.~L.}\ \bibnamefont {Gardel}}, \ and\
  \bibinfo {author} {\bibfnamefont {S.~R.}\ \bibnamefont {Nagel}},\ }\href@noop
  {} {\bibfield  {journal} {\bibinfo  {journal} {Phys. Rev. E}\ }\textbf
  {\bibinfo {volume} {87}},\ \bibinfo {pages} {012303} (\bibinfo {year}
  {2013})}\BibitemShut {NoStop}%
\bibitem [{\citenamefont {Royall}\ \emph {et~al.}(2008)\citenamefont {Royall},
  \citenamefont {Williams}, \citenamefont {Ohtsuka},\ and\ \citenamefont
  {Tanaka}}]{royall08}%
  \BibitemOpen
  \bibfield  {author} {\bibinfo {author} {\bibfnamefont {C.~P.}\ \bibnamefont
  {Royall}}, \bibinfo {author} {\bibfnamefont {S.~R.}\ \bibnamefont
  {Williams}}, \bibinfo {author} {\bibfnamefont {T.}~\bibnamefont {Ohtsuka}}, \
  and\ \bibinfo {author} {\bibfnamefont {H.}~\bibnamefont {Tanaka}},\
  }\href@noop {} {\bibfield  {journal} {\bibinfo  {journal} {Nature Mat.}\
  }\textbf {\bibinfo {volume} {7}},\ \bibinfo {pages} {556} (\bibinfo {year}
  {2008})}\BibitemShut {NoStop}%
\bibitem [{\citenamefont {Royall}\ and\ \citenamefont
  {Williams}(2015)}]{royall15}%
  \BibitemOpen
  \bibfield  {author} {\bibinfo {author} {\bibfnamefont {C.~P.}\ \bibnamefont
  {Royall}}\ and\ \bibinfo {author} {\bibfnamefont {S.~R.}\ \bibnamefont
  {Williams}},\ }\href@noop {} {\bibfield  {journal} {\bibinfo  {journal}
  {Phys. Rep.}\ }\textbf {\bibinfo {volume} {560}},\ \bibinfo {pages} {1}
  (\bibinfo {year} {2015})}\BibitemShut {NoStop}%
\bibitem [{\citenamefont {Maxwell}(1864)}]{maxwell64}%
  \BibitemOpen
  \bibfield  {author} {\bibinfo {author} {\bibfnamefont {J.~C.}\ \bibnamefont
  {Maxwell}},\ }\href@noop {} {\bibfield  {journal} {\bibinfo  {journal}
  {Philos. Mag. (4th ser.)}\ }\textbf {\bibinfo {volume} {27}},\ \bibinfo
  {pages} {250} (\bibinfo {year} {1864})}\BibitemShut {NoStop}%
\bibitem [{\citenamefont {O'Hern}\ \emph {et~al.}(2003)\citenamefont {O'Hern},
  \citenamefont {Silbert}, \citenamefont {Liu},\ and\ \citenamefont
  {Nagel}}]{ohern02}%
  \BibitemOpen
  \bibfield  {author} {\bibinfo {author} {\bibfnamefont {C.~S.}\ \bibnamefont
  {O'Hern}}, \bibinfo {author} {\bibfnamefont {L.~E.}\ \bibnamefont {Silbert}},
  \bibinfo {author} {\bibfnamefont {A.~J.}\ \bibnamefont {Liu}}, \ and\
  \bibinfo {author} {\bibfnamefont {S.~R.}\ \bibnamefont {Nagel}},\ }\href@noop
  {} {\bibfield  {journal} {\bibinfo  {journal} {Phys. Rev. E}\ }\textbf
  {\bibinfo {volume} {68}},\ \bibinfo {pages} {011306} (\bibinfo {year}
  {2003})}\BibitemShut {NoStop}%
\bibitem [{\citenamefont {Karayiannis}\ and\ \citenamefont
  {Laso}(2008)}]{karayiannis08}%
  \BibitemOpen
  \bibfield  {author} {\bibinfo {author} {\bibfnamefont {N.~C.}\ \bibnamefont
  {Karayiannis}}\ and\ \bibinfo {author} {\bibfnamefont {M.}~\bibnamefont
  {Laso}},\ }\href@noop {} {\bibfield  {journal} {\bibinfo  {journal} {Phys.
  Rev. Lett.}\ }\textbf {\bibinfo {volume} {100}},\ \bibinfo {pages} {050602}
  (\bibinfo {year} {2008})}\BibitemShut {NoStop}%
\bibitem [{\citenamefont {Hoy}\ and\ \citenamefont {O'Hern}(2012)}]{hoy12}%
  \BibitemOpen
  \bibfield  {author} {\bibinfo {author} {\bibfnamefont {R.~S.}\ \bibnamefont
  {Hoy}}\ and\ \bibinfo {author} {\bibfnamefont {C.~S.}\ \bibnamefont
  {O'Hern}},\ }\href@noop {} {\bibfield  {journal} {\bibinfo  {journal} {Soft
  Matter}\ }\textbf {\bibinfo {volume} {8}},\ \bibinfo {pages} {1215} (\bibinfo
  {year} {2012})}\BibitemShut {NoStop}%
\bibitem [{\citenamefont {Lappala}\ \emph {et~al.}(2016)\citenamefont
  {Lappala}, \citenamefont {Zaccone},\ and\ \citenamefont
  {Terentjev}}]{lappala16}%
  \BibitemOpen
  \bibfield  {author} {\bibinfo {author} {\bibfnamefont {A.}~\bibnamefont
  {Lappala}}, \bibinfo {author} {\bibfnamefont {A.}~\bibnamefont {Zaccone}}, \
  and\ \bibinfo {author} {\bibfnamefont {E.~M.}\ \bibnamefont {Terentjev}},\
  }\href@noop {} {\bibfield  {journal} {\bibinfo  {journal} {Soft Matter}\
  }\textbf {\bibinfo {volume} {12}},\ \bibinfo {pages} {7330} (\bibinfo {year}
  {2016})}\BibitemShut {NoStop}%
\bibitem [{\citenamefont {Zaccone}\ and\ \citenamefont
  {Terentjev}(2013)}]{zaccone13}%
  \BibitemOpen
  \bibfield  {author} {\bibinfo {author} {\bibfnamefont {A.}~\bibnamefont
  {Zaccone}}\ and\ \bibinfo {author} {\bibfnamefont {E.~M.}\ \bibnamefont
  {Terentjev}},\ }\href@noop {} {\bibfield  {journal} {\bibinfo  {journal}
  {Phys. Rev. Lett.}\ }\textbf {\bibinfo {volume} {110}},\ \bibinfo {pages}
  {178002} (\bibinfo {year} {2013})}\BibitemShut {NoStop}%
\bibitem [{\citenamefont {Ballauf}(1989)}]{ballauff89}%
  \BibitemOpen
  \bibfield  {author} {\bibinfo {author} {\bibfnamefont {M.}~\bibnamefont
  {Ballauf}},\ }\href@noop {} {\bibfield  {journal} {\bibinfo  {journal}
  {Angew. Chem.}\ }\textbf {\bibinfo {volume} {28}},\ \bibinfo {pages} {253}
  (\bibinfo {year} {1989})}\BibitemShut {NoStop}%
\bibitem [{\citenamefont {Strobl}(2007)}]{strobl07}%
  \BibitemOpen
  \bibfield  {author} {\bibinfo {author} {\bibfnamefont {G.}~\bibnamefont
  {Strobl}},\ }\href@noop {} {\emph {\bibinfo {title} {The Physics of
  Polymers}}}\ (\bibinfo  {publisher} {Springer},\ \bibinfo {year}
  {2007})\BibitemShut {NoStop}%
\bibitem [{\citenamefont {Nguyen}\ \emph {et~al.}(2015)\citenamefont {Nguyen},
  \citenamefont {Smith}, \citenamefont {Hoy},\ and\ \citenamefont
  {Karayiannis}}]{nguyen15}%
  \BibitemOpen
  \bibfield  {author} {\bibinfo {author} {\bibfnamefont {H.~T.}\ \bibnamefont
  {Nguyen}}, \bibinfo {author} {\bibfnamefont {T.~B.}\ \bibnamefont {Smith}},
  \bibinfo {author} {\bibfnamefont {R.~S.}\ \bibnamefont {Hoy}}, \ and\
  \bibinfo {author} {\bibfnamefont {N.~C.}\ \bibnamefont {Karayiannis}},\
  }\href@noop {} {\bibfield  {journal} {\bibinfo  {journal} {J. Chem. Phys.}\
  }\textbf {\bibinfo {volume} {143}},\ \bibinfo {pages} {144901} (\bibinfo
  {year} {2015})}\BibitemShut {NoStop}%
\bibitem [{\citenamefont {Zou}\ \emph {et~al.}(2009)\citenamefont {Zou},
  \citenamefont {Cheng}, \citenamefont {Rivers}, \citenamefont {Jaeger},\ and\
  \citenamefont {Nagel}}]{zou09}%
  \BibitemOpen
  \bibfield  {author} {\bibinfo {author} {\bibfnamefont {L.-N.}\ \bibnamefont
  {Zou}}, \bibinfo {author} {\bibfnamefont {X.}~\bibnamefont {Cheng}}, \bibinfo
  {author} {\bibfnamefont {M.~L.}\ \bibnamefont {Rivers}}, \bibinfo {author}
  {\bibfnamefont {H.~M.}\ \bibnamefont {Jaeger}}, \ and\ \bibinfo {author}
  {\bibfnamefont {S.~R.}\ \bibnamefont {Nagel}},\ }\href@noop {} {\bibfield
  {journal} {\bibinfo  {journal} {Science}\ }\textbf {\bibinfo {volume}
  {326}},\ \bibinfo {pages} {408} (\bibinfo {year} {2009})}\BibitemShut
  {NoStop}%
\bibitem [{\citenamefont {Hoy}(2017)}]{hoy17}%
  \BibitemOpen
  \bibfield  {author} {\bibinfo {author} {\bibfnamefont {R.~S.}\ \bibnamefont
  {Hoy}},\ }\href@noop {} {\bibfield  {journal} {\bibinfo  {journal} {Phys.
  Rev. Lett.}\ }\textbf {\bibinfo {volume} {118}},\ \bibinfo {pages} {068002}
  (\bibinfo {year} {2017})}\BibitemShut {NoStop}%
\bibitem [{\citenamefont {Hoy}\ and\ \citenamefont
  {Karayiannis}(2013)}]{hoy13}%
  \BibitemOpen
  \bibfield  {author} {\bibinfo {author} {\bibfnamefont {R.~S.}\ \bibnamefont
  {Hoy}}\ and\ \bibinfo {author} {\bibfnamefont {N.~C.}\ \bibnamefont
  {Karayiannis}},\ }\href@noop {} {\bibfield  {journal} {\bibinfo  {journal}
  {Phys. Rev. E}\ }\textbf {\bibinfo {volume} {88}},\ \bibinfo {pages} {012601}
  (\bibinfo {year} {2013})}\BibitemShut {NoStop}%
\bibitem [{\citenamefont {Phillips}(1979)}]{phillips79}%
  \BibitemOpen
  \bibfield  {author} {\bibinfo {author} {\bibfnamefont {J.~C.}\ \bibnamefont
  {Phillips}},\ }\href@noop {} {\bibfield  {journal} {\bibinfo  {journal} {J.
  Non-Cryst. Solids}\ }\textbf {\bibinfo {volume} {34}},\ \bibinfo {pages}
  {153} (\bibinfo {year} {1979})}\BibitemShut {NoStop}%
\bibitem [{\citenamefont {Thorpe}(1983)}]{thorpe83}%
  \BibitemOpen
  \bibfield  {author} {\bibinfo {author} {\bibfnamefont {M.~F.}\ \bibnamefont
  {Thorpe}},\ }\href@noop {} {\bibfield  {journal} {\bibinfo  {journal} {J.
  Non-Cryst. Solids}\ }\textbf {\bibinfo {volume} {57}},\ \bibinfo {pages}
  {355} (\bibinfo {year} {1983})}\BibitemShut {NoStop}%
\bibitem [{\citenamefont {Micoulaut}(2016)}]{micoulaut16}%
  \BibitemOpen
  \bibfield  {author} {\bibinfo {author} {\bibfnamefont {M.}~\bibnamefont
  {Micoulaut}},\ }\href@noop {} {\bibfield  {journal} {\bibinfo  {journal}
  {Adv. Phys. X}\ }\textbf {\bibinfo {volume} {1}},\ \bibinfo {pages} {147}
  (\bibinfo {year} {2016})}\BibitemShut {NoStop}%
\bibitem [{\citenamefont {Kremer}\ and\ \citenamefont
  {Grest}(1990)}]{kremer90}%
  \BibitemOpen
  \bibfield  {author} {\bibinfo {author} {\bibfnamefont {K.}~\bibnamefont
  {Kremer}}\ and\ \bibinfo {author} {\bibfnamefont {G.~S.}\ \bibnamefont
  {Grest}},\ }\href@noop {} {\bibfield  {journal} {\bibinfo  {journal} {J.
  Chem. Phys.}\ }\textbf {\bibinfo {volume} {92}},\ \bibinfo {pages} {5057}
  (\bibinfo {year} {1990})}\BibitemShut {NoStop}%
\bibitem [{\citenamefont {Auhl}\ \emph {et~al.}(2003)\citenamefont {Auhl},
  \citenamefont {Everarers}, \citenamefont {Grest}, \citenamefont {Kremer},\
  and\ \citenamefont {Plimpton}}]{auhl03}%
  \BibitemOpen
  \bibfield  {author} {\bibinfo {author} {\bibfnamefont {R.}~\bibnamefont
  {Auhl}}, \bibinfo {author} {\bibfnamefont {R.}~\bibnamefont {Everarers}},
  \bibinfo {author} {\bibfnamefont {G.~S.}\ \bibnamefont {Grest}}, \bibinfo
  {author} {\bibfnamefont {K.}~\bibnamefont {Kremer}}, \ and\ \bibinfo {author}
  {\bibfnamefont {S.~J.}\ \bibnamefont {Plimpton}},\ }\href@noop {} {\bibfield
  {journal} {\bibinfo  {journal} {J. Chem. Phys.}\ }\textbf {\bibinfo {volume}
  {119}},\ \bibinfo {pages} {12718} (\bibinfo {year} {2003})}\BibitemShut
  {NoStop}%
\bibitem [{\citenamefont {Nguyen}\ and\ \citenamefont {Hoy}(2016)}]{nguyen16}%
  \BibitemOpen
  \bibfield  {author} {\bibinfo {author} {\bibfnamefont {H.~T.}\ \bibnamefont
  {Nguyen}}\ and\ \bibinfo {author} {\bibfnamefont {R.~S.}\ \bibnamefont
  {Hoy}},\ }\href@noop {} {\bibfield  {journal} {\bibinfo  {journal} {Phys.
  Rev. E}\ }\textbf {\bibinfo {volume} {94}},\ \bibinfo {pages} {052502}
  (\bibinfo {year} {2016})}\BibitemShut {NoStop}%
\bibitem [{\citenamefont {Plimpton}(1995)}]{plimpton95}%
  \BibitemOpen
  \bibfield  {author} {\bibinfo {author} {\bibfnamefont {S.}~\bibnamefont
  {Plimpton}},\ }\href@noop {} {\bibfield  {journal} {\bibinfo  {journal} {J.
  Comp. Phys.}\ }\textbf {\bibinfo {volume} {117}},\ \bibinfo {pages} {1}
  (\bibinfo {year} {1995})}\BibitemShut {NoStop}%
\bibitem [{Note1()}]{Note1}%
  \BibitemOpen
  \bibinfo {note} {As in Ref.\ \cite {nguyen15}, values of $T_s$ were
  determined by locating the jump in packing fraction $\phi (T)$ for
  crystallizing systems, or the intersection of low-$T$ and high-$T$ linear
  fits to $\phi (T)$ for glassforming systems.}\BibitemShut {Stop}%
\bibitem [{\citenamefont {Foteinopoulou}\ \emph {et~al.}(2008)\citenamefont
  {Foteinopoulou}, \citenamefont {Karayiannis}, \citenamefont {Laso},
  \citenamefont {Kr{\"o}ger},\ and\ \citenamefont
  {Mansfield}}]{foteinopoulou08}%
  \BibitemOpen
  \bibfield  {author} {\bibinfo {author} {\bibfnamefont {K.}~\bibnamefont
  {Foteinopoulou}}, \bibinfo {author} {\bibfnamefont {N.~C.}\ \bibnamefont
  {Karayiannis}}, \bibinfo {author} {\bibfnamefont {M.}~\bibnamefont {Laso}},
  \bibinfo {author} {\bibfnamefont {M.}~\bibnamefont {Kr{\"o}ger}}, \ and\
  \bibinfo {author} {\bibfnamefont {M.~L.}\ \bibnamefont {Mansfield}},\
  }\href@noop {} {\bibfield  {journal} {\bibinfo  {journal} {Phys. Rev. Lett.}\
  }\textbf {\bibinfo {volume} {101}},\ \bibinfo {pages} {265702} (\bibinfo
  {year} {2008})}\BibitemShut {NoStop}%
\bibitem [{\citenamefont {Kyrylyuk}\ and\ \citenamefont
  {Philipse}(2011)}]{kyrylyuk11}%
  \BibitemOpen
  \bibfield  {author} {\bibinfo {author} {\bibfnamefont {A.~V.}\ \bibnamefont
  {Kyrylyuk}}\ and\ \bibinfo {author} {\bibfnamefont {A.~P.}\ \bibnamefont
  {Philipse}},\ }\href@noop {} {\bibfield  {journal} {\bibinfo  {journal}
  {Phys. Status Solidi A}\ }\textbf {\bibinfo {volume} {208}},\ \bibinfo
  {pages} {2299} (\bibinfo {year} {2011})}\BibitemShut {NoStop}%
\bibitem [{\citenamefont {Broedersz}\ \emph {et~al.}(2011)\citenamefont
  {Broedersz}, \citenamefont {Mao}, \citenamefont {Lubensky},\ and\
  \citenamefont {Mackintosh}}]{broedersz11}%
  \BibitemOpen
  \bibfield  {author} {\bibinfo {author} {\bibfnamefont {C.~P.}\ \bibnamefont
  {Broedersz}}, \bibinfo {author} {\bibfnamefont {X.}~\bibnamefont {Mao}},
  \bibinfo {author} {\bibfnamefont {T.~C.}\ \bibnamefont {Lubensky}}, \ and\
  \bibinfo {author} {\bibfnamefont {F.~C.}\ \bibnamefont {Mackintosh}},\
  }\href@noop {} {\bibfield  {journal} {\bibinfo  {journal} {Nature Phys.}\
  }\textbf {\bibinfo {volume} {7}},\ \bibinfo {pages} {983} (\bibinfo {year}
  {2011})}\BibitemShut {NoStop}%
\bibitem [{\citenamefont {Lois}\ and\ \citenamefont {O'Hern}(2008)}]{lois08}%
  \BibitemOpen
  \bibfield  {author} {\bibinfo {author} {\bibfnamefont {G.}~\bibnamefont
  {Lois}}\ and\ \bibinfo {author} {\bibfnamefont {C.~S.}\ \bibnamefont
  {O'Hern}},\ }\href@noop {} {\bibfield  {journal} {\bibinfo  {journal} {Phys.
  Rev. Lett.}\ }\textbf {\bibinfo {volume} {100}},\ \bibinfo {pages} {028001}
  (\bibinfo {year} {2008})}\BibitemShut {NoStop}%
\bibitem [{\citenamefont {Berthier}\ and\ \citenamefont
  {Tarjus}(2009)}]{berthier09}%
  \BibitemOpen
  \bibfield  {author} {\bibinfo {author} {\bibfnamefont {L.}~\bibnamefont
  {Berthier}}\ and\ \bibinfo {author} {\bibfnamefont {G.}~\bibnamefont
  {Tarjus}},\ }\href@noop {} {\bibfield  {journal} {\bibinfo  {journal} {Phys.
  Rev. Lett.}\ }\textbf {\bibinfo {volume} {102}},\ \bibinfo {pages} {170601}
  (\bibinfo {year} {2009})}\BibitemShut {NoStop}%
\bibitem [{\citenamefont {Schreck}\ \emph {et~al.}(2010)\citenamefont
  {Schreck}, \citenamefont {Xu},\ and\ \citenamefont {O'Hern}}]{schreck10}%
  \BibitemOpen
  \bibfield  {author} {\bibinfo {author} {\bibfnamefont {C.~F.}\ \bibnamefont
  {Schreck}}, \bibinfo {author} {\bibfnamefont {N.}~\bibnamefont {Xu}}, \ and\
  \bibinfo {author} {\bibfnamefont {C.~S.}\ \bibnamefont {O'Hern}},\
  }\href@noop {} {\bibfield  {journal} {\bibinfo  {journal} {Soft Matt.}\
  }\textbf {\bibinfo {volume} {6}},\ \bibinfo {pages} {2960} (\bibinfo {year}
  {2010})}\BibitemShut {NoStop}%
\bibitem [{\citenamefont {Liu}\ and\ \citenamefont {Nagel}(1998)}]{liu98}%
  \BibitemOpen
  \bibfield  {author} {\bibinfo {author} {\bibfnamefont {A.~J.}\ \bibnamefont
  {Liu}}\ and\ \bibinfo {author} {\bibfnamefont {S.~R.}\ \bibnamefont
  {Nagel}},\ }\href@noop {} {\bibfield  {journal} {\bibinfo  {journal}
  {Nature}\ }\textbf {\bibinfo {volume} {396}},\ \bibinfo {pages} {21}
  (\bibinfo {year} {1998})}\BibitemShut {NoStop}%
\end{thebibliography}

%

\end{document}